\documentclass[12pt]{article}

\usepackage{amsmath, psfig, amsfonts}
\usepackage{graphics}
\usepackage[dvips]{graphicx}
\usepackage{epsfig}
\usepackage{subfigure}
\usepackage[dvips]{color}
 \usepackage{amsfonts}
\usepackage{latexsym}
\usepackage{amsmath,amssymb}
\usepackage{fancyhdr,cite}
\usepackage{psfrag}
\usepackage{cite}

\newtheorem{theorem}{Theorem}

\DeclareMathOperator*{\E}{E}    

\begin{document}
\renewcommand{\textfraction}{0}
\title{A Capacity Achieving and Low Complexity  Multilevel Coding Scheme for ISI Channels}
\author{\normalsize
Mei Chen, Teng Li and Oliver M. Collins\\
\small Dept. of Electrical Engineering \\[-5pt] \small  University of Notre Dame\\
[-5pt]\small Notre Dame,  IN 46556 \\
[-5pt] \small \{mchen1, tli, ocollins\}@nd.edu}

\date{}
\maketitle

\thispagestyle{empty}
\begin{abstract}
We propose a computationally efficient multilevel coding scheme to
achieve the capacity of an ISI channel using layers of binary
inputs. The transmitter employs multilevel coding with linear
mapping. The receiver uses multistage decoding where each stage
performs a separate linear minimum mean square error (LMMSE)
equalization and decoding.  The optimality of the scheme is due to
the fact that the LMMSE equalizer is information lossless in an
ISI channel when signal to noise ratio is sufficiently low. The
computational complexity is low and scales linearly with the
length of the channel impulse response and the number of layers.
The decoder at each layer sees an equivalent AWGN channel, which
makes coding straightforward.
\end{abstract}
\normalsize

\section{introduction}

High bandwidth efficient communication systems require the use of
multilevel or multi-phase constellations. The major difficulty of
applying a coded modulation scheme, such as trellis coded
modulation  \cite{underboeckIT82} and  bit-interleaved coded
modulation  \cite{caireIT98}, to an intersymbol interference (ISI)
channel is receiver complexity. Clearly, the optimal joint
equalization and decoding scheme, such as \cite{chevillatCOM89},
leads to extremely high complexity. Even in the context of Turbo
equalization \cite{kotterCom02}, a suboptimal yet very efficient
iterative approach, the optimal BCJR-based equalizer has
exponential complexity in both channel memory and constellation
size. Various equalizers with reduced complexity, such as minimum
mean square error (MMSE) equalizer, are proposed in
\cite{dejongheICC2002}, \cite{magniezAsilomar2000} at a cost of
degraded performance.

This paper proposes a coded modulation scheme for a static ISI
channel with capacity achieving performance and very efficient
computation.  The idea is to do multilevel coding (MLC)
\cite{imaiIT77}, \cite{wachsmannIT99} and multistage decoding with
a linear mapping \cite{duanISIT97} at the transmitter and a
separate linear MMSE (LMMSE) equalization and decoding at the
receiver. Minimal computation is required and will scale linearly
with channel length and the number of layers.  The number of
layers $M$ is large so that, by the central limiting theorem, the
output of the mapper approaches a Gaussian distribution. More
importantly,   this paper shows that an LMMSE equalizer in an ISI
channel with non-white Gaussian noise is information lossless when
the signal to noise ratio (SNR) is sufficiently low. Therefore,
given a fixed total power, a large $M$ will effectively  drive
each layer to operate at a low SNR region, in which, the proposed
separate LMMSE equalization and decoding is not only efficient but
also optimal. The task of code design is also simple since the
decoder at each layer sees an equivalent AWGN channel.
Consequently, the scheme can achieve the i.i.d. Gaussian capacity
of an ISI channel if each layer employs a capacity achieving code.

This paper is organized as follows. Section \ref{sec:system-model}
presents the MLC scheme with the LMMSE equalization algorithm. The
optimality of the scheme is discussed analytically in Section
\ref{sec:optimality}, and also is  numerically demonstrated in
Section \ref{sec:design-issue}. Some design issues, such as power
allocation and rate design, are discussed in Section
\ref{sec:design-issue}. The result shows that if the power of each
level is properly allocated,  only a moderate number of layers is
required for the overall system to approach the channel capacity.

\section{System Model}\label{sec:system-model}
\begin{figure}[t]
\psfrag{x1}{$x_1(k)$}
\psfrag{x2}{$x_2(k)$}
\psfrag{xM}{$x_M(k)$}
\psfrag{xhat1}{$\widehat{x}_1(k)$}
\psfrag{xhat2}{$\widehat{x}_2(k)$}
\psfrag{xhatM}{$\widehat{x}_M(k)$}
\psfrag{xhatM-1}{$\widehat{x}_{M-1}(k)$}
\psfrag{x(k)}{$x(k)$}
\psfrag{y(h)}{$y(k)$}
\psfrag{h}{$h$}
\psfrag{w(k)}{$w(k)$}
\centering
\includegraphics[width =6in]{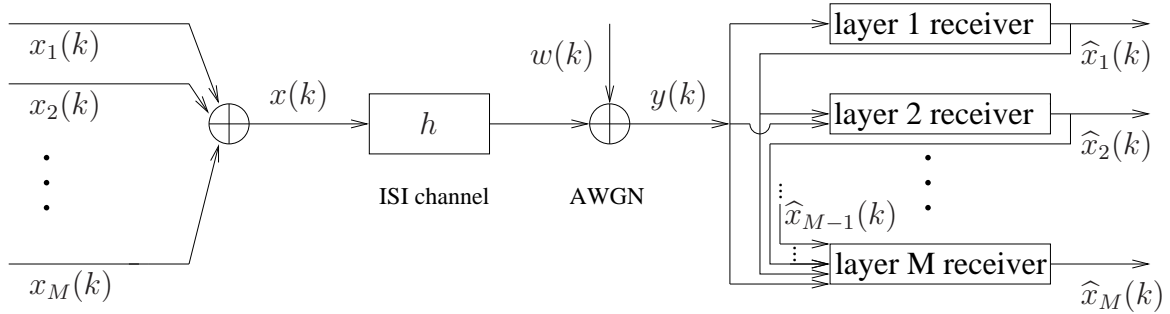}
      \caption{System model.}
      \label{fig:channel_model}
\end{figure}

This section presents the overall system of the  MLC with LMMSE
equalization as shown in Fig. \ref{fig:channel_model}. Let $\{
x_j(k)\}_{k=1}^N$ be the scaled version of a sequence of
identically, independently and uniformly distributed binary
sequences, such that $ x_j(k) = -\!\sqrt{P_j}$ or $ +\!\sqrt{P_j}$
with equal probability for all $k$.  A total of $M$ such mutually
independent BPSK sequences, $\{ x_j(k)\}_{k=1}^N$ for $j =
1,\cdots, M$,  are summed to produce the transmitted sequence $\{
x(k)\}_{k=1}^N$, where $x(k) = \sum_{j=1}^Mx_j(k)$.  In general,
each  BPSK layer has a different power level $P_j$ and the total
power is $P = \sum_{j=1}^MP_j$. The ISI channel is modelled as a
time-invariant causal FIR filter with order $L_h$ and an impulse
response $h = \{ h(0),\cdots, h(L_h)\}$ that is known to both the
transmitter and the receiver. Let $w(k)\sim \mathcal{N}(0,
\sigma_w^2)$ be the AWGN. The received signal is written as
\begin{equation}
y(k) = \sum_{j = 1}^{M}\sum_{i=0}^{L_h}h(i)x_j(k-i) + w(k).
\end{equation}

\begin{figure}[t]
\psfrag{1}{$y(k)$} \psfrag{3}{$\widetilde{y}_m(k)$}
\psfrag{4}{$\widetilde{x}_m(k)$} \psfrag{5}{APP}
\psfrag{2}{$\sum\limits_{j=1}^{m-1}\sum\limits_{i=0}^{L_h}h(i)\widehat{x}_j(k-1)$}
\psfrag{6}{$\widehat{x}_m(k)$} \centering
\includegraphics[width =6in]{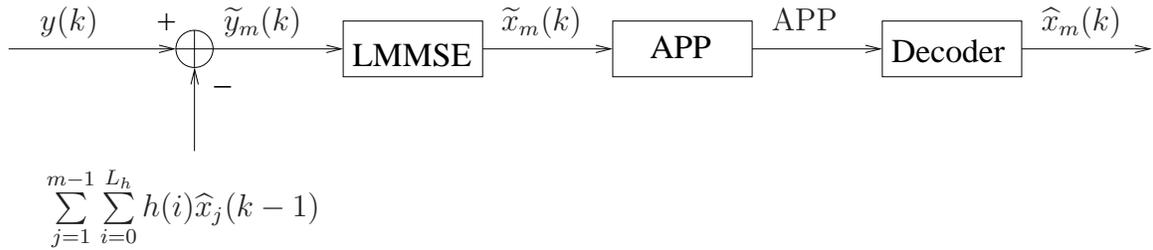}
      \caption{Layer $m$ receiver.}
      \label{fig:receiver}
\end{figure}

The receiver employs multistage decoding, i.e., it successively
equalizes and decodes $M$ BPSK layers from layer 1 to layer $M$.
Once a layer is correctly decoded, its effect on the channel
output can be completely removed by filtering it with $h$ and
subtracting the resulting waveform from the received signal. After
cancelling the interferences of all decoded layers, the current
layer treats the undecoded layers as noise and  employs  an LMMSE
filter to compute the likelihood ratio of the input bit. A
memoryless channel decoder uses these likelihood ratios to decode
the codeword and feed the hard decision to the next layer. This
receiver of layer $m$ is shown in Fig \ref{fig:receiver}.

Consider the decoding process of the $m$th stage. We assume the
receiver has correctly decoded layer 1 to $m-1$, i.e.
$\widehat{x}_j(k) = x_j(k)$ for all $k$ and $j=1,\cdots, m-1$. The
decoded layer interference cancelled signal at layer $m$ is
\begin{equation}
\widetilde{y}_m(k) = \sum_{j = m}^{M}\sum_{i=0}^{L_h}h(i)x_j(k-i)
+ w(k).
\end{equation}
The receiver uses an FIR LMMSE filter of order $2L_g$
 to estimate $x_m(k)$ from a vector of channel output $\widetilde{\mathbf{y}}_m
(k) \triangleq [\widetilde{y}_m (k-L_g),\cdots,\widetilde{y}_m
(k+L_g)]^T$ for any $k$. Using vector notation, define a Toeplitz
matrix $\mathbf{H}\in\mathcal{R}^{(2L_g+1) \times (2L_g + L_h) }$
and its block partition  as
\begin{equation}\label{eqn:H}
\mathbf{H} = \left[
\begin{array}{ccccccc}
h(L_h)        & \cdots    & h(0)  &0  &\cdots &\cdots & 0\\
0    &h(L_h)      &\cdots &h(0)   & 0 &\cdots &0 \\
\vdots          &\ddots     &\ddots  &\ddots  &\ddots &\ddots  &\vdots   \\
0   &\cdots     &0     &0 &h(L_h)     &\cdots &\ h(0) \\
\end{array}
\right]= [\mathbf{H}_1, {\bf h}_k, \mathbf{H}_2]
\end{equation}
 where
$\mathbf{h}_k\in\mathcal{R}^{(2L_g+1) \times 1 } $,
$\mathbf{H}_1\in\mathcal{R}^{(2L_g+1) \times (L_g + L_h) } $
 and $\mathbf{H}_2\in\mathcal{R}^{(2L_g+1) \times L_g } $. Furthermore,
define $ \mathbf{H}^{isi}= [\mathbf{H}_1, \mathbf{H}_2]$.
Therefore,
\begin{equation}\label{eqn:vec-channel}
\widetilde{\mathbf{y}}_m (k)= \mathbf{h}_kx_m(k) +
\mathbf{H}^{isi}\mathbf{x}^{isi}_m (k) + \sum_{j =
m+1}^{M}\mathbf{Hx}_j(k) + \mathbf{w}(k),
\end{equation}
where $\mathbf{x}_j(k)\triangleq [x_j(k-L_g),\cdots, x_j(k+L_g)]^T
$, $\mathbf{x}^{isi}_m (k) \triangleq [x_m(k-L_g), \cdots,
x_m(k-1), x_m(k+1),\cdots, x_m(k+L_g)]^T$ and
$\mathbf{w}(k)\triangleq [w(k-L_g),\cdots, w(k+L_g)]^T $. In
\eqref{eqn:vec-channel}, the first term is due to the desired
input, the second term is the ISI, the third term is the
interference from undecoded layers and the last term is AWGN. From
linear estimation theory, the LMMSE filter at layer $m$ in vector
form is
\begin{equation}\label{eqn:LMMSE-filter}
\mathbf{g}_m =
\E[\widetilde{\mathbf{y}}_m(k)\widetilde{\mathbf{y}}_m^\dag(k)
]^{-1}\E[\widetilde{\mathbf{y}}_mx^\dag_m(k) ] = P_m
\Big(\sum_{j=m}^MP_j \mathbf{HH}^\dag + \sigma_w^2
\mathbf{I}_{2L_g+1}\Big)^{-1}\mathbf{h}_k.
\end{equation}
The LMMSE estimate of $x_m(k)$ is thus
\begin{equation}\label{eqn:lmmse-x}
\widetilde{x}_m(k) = \mathbf{g}_m^\dag\widetilde{\mathbf{y}}_m
(k).
\end{equation}
From \eqref{eqn:vec-channel}, \eqref{eqn:lmmse-x} can be
equivalently written as
\begin{equation}\label{eqn:eqv-chan}
\widetilde{x}_m(k) =  \alpha_m x_m(k) + \zeta_m(k),
\end{equation}
where $\alpha_m=  \mathbf{g}_m^\dag  \mathbf{h}_k$ is the
equivalent channel gain and
\begin{equation}
\zeta_m(k)= \mathbf{g}_m^\dag\mathbf{H}^{isi}\mathbf{x}^{isi}_m
(k) + \sum_{j = m+1}^{M} \mathbf{g}_m^\dag\mathbf{Hx}_j(k) +
\mathbf{g}_m^\dag\mathbf{w}(k)
\end{equation} is the equivalent
channel noise. From the central limiting theorem, the equivalent
noise has a Gaussian distribution such that $\zeta_m(k) \sim
\mathcal{N}(0,\sigma_{\zeta_m}^2)$, where
\begin{equation}
\sigma_{\zeta_m}^2 =
P_m\mathbf{g}_m^\dag\mathbf{H}^{isi}(\mathbf{H}^{isi})^\dag
\mathbf{g}_m +  \sum_{j = m+1}^{M}P_j
\mathbf{g}_m^\dag\mathbf{H}\mathbf{H}^\dag\mathbf{g}_m
+\sigma_w^2\mathbf{g}^\dag\mathbf{g}.
\end{equation}
Thus, the
likelihood function of $x_m(k)$  is computed from
\begin{align}\label{eqn:likelihood}
&P\big(\widetilde{x}_m(k) | {x}_m(k) \big)
=\frac{1}{\sqrt{2\pi\sigma^2_{\zeta_m}}}
\exp\bigg(-\frac{\big(\widetilde{x}_m(k)- \alpha_m
 {x}_m(k)\big)^2}{2\sigma^2_{\zeta_m}}\bigg).
\end{align}
The subsequent decoder decodes the codeword solely based on the
set of likelihood ratios. Its hard decision is feedback to the
next stage.

\section{Optimality}\label{sec:optimality}

In this section, Theorem \ref{thm:1} and \ref{thm:main} will
establish that given a sufficiently large number of layers, the
proposed multilevel coding and its receiver, which sequentially
(non-iteratively) performs  LMMSE equalization and decoding, can
approach the ISI channel capacity with  i.i.d. Gaussian input.

The key observation  is that, at any layer, if the signal to
interference plus noise ratio (SINR)  is sufficiently low, the
LMMSE filtered channel retains the original ISI channel capacity.
Theorem \ref{thm:1} shows this   for Gaussian input and Gaussian
interference. Theorem \ref{thm:main} extends this results to the
$M$-layered MLC scheme with  BPSK inputs.  Throughout this paper,
the vector notation ${\bf x} = [x(0), \cdots, x(N-1)]^T$ will be
used.

\begin{theorem}\label{thm:1}
Consider an ISI channel with input $x(k)$,  interference $z(k)$
and noise $w(k)$,
\begin{equation}\label{eqn:isichannel}
    y(k) = \sum_{i=0}^{L_h}h(i)\big(x(k-i)+z(k-i)\big)+w(k),
\end{equation}
where $h=\{ h(0),\cdots,h(L_h)\}$ is the channel impulse response,
and $x(k)$, $z(k)$ and $w(k)$ are i.i.d. Gaussian random sequences
with zero mean and variance $\sigma_x^2$, $\sigma_z^2$ and
$\sigma_w^2$ respectively. Let $\widetilde{x}(k)$ be the output of
the LMMSE filter, then
\begin{equation}
    \lim_{\sigma_x^2/(\sigma_z^2 + \sigma_w^2)\rightarrow 0}
\frac{I(\mathbf{x}; \widetilde{\mathbf{x}})}
{I(\mathbf{x}; \mathbf{y})} = 1. \label{eqn:theorem1}
\end{equation}
\end{theorem}

\begin{theorem}\label{thm:main}
Let $C^{g,ISI}$ be the ISI channel capacity with Gaussian inputs.
Let $C^{b,MLC}$ be the capacity of proposed $M$-layer multilevel coding with LMMSE equalization, where the input of each layer is i.i.d. BPSK of power $P_m$ and the LMMSE filter has  infinite length.   If the  total power $P$ is finite  and the power allocation at each  layer satisfies
\begin{equation}
P_m \ll (\sigma_w^2 + \sum_{j=m+1}^{M} P_j) \quad \text{ for $m =
1,\cdots, M$ }
\end{equation}
where $\sigma_w^2$ is the
variance of AWGN,  then
  \begin{equation}\label{eqn:thm2-result}
    \lim_{M\to\infty}C^{b,MLC} = C^{g,ISI}.
  \end{equation}
\end{theorem}

The proof of Theorem \ref{thm:1} and \ref{thm:main} is omitted due
to space limitation.

\begin{figure}[t]
\centering
\includegraphics[width =4.7in]{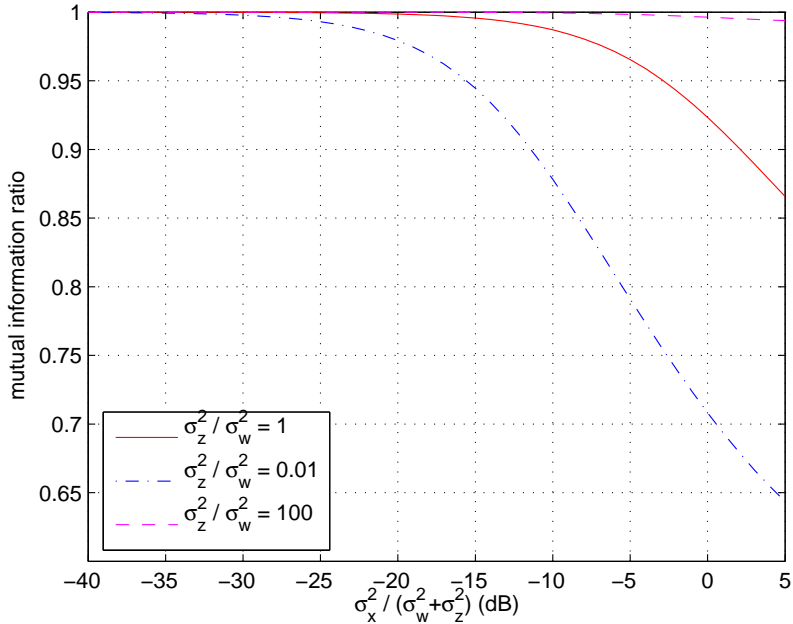}
\caption{Convergence of  the ratio of LMMSE filtered capacity over
the original ISI channel capacity as SINR decreases for Gaussian
input and interference.}
      \label{fig:convergence}
\end{figure}

The convergence of \eqref{eqn:theorem1} as ${\sigma_x^2/
(\sigma_z^2 + \sigma_w^2)\rightarrow 0}$ is illustrated in Fig.
\ref{fig:convergence} for different values of
${\sigma_z^2}/{\sigma_w^2}$ and a random  $10$-tap ISI channel.
The LMMSE used is a $401$-tap FIR filter as designed according to
\eqref{eqn:LMMSE-filter}.  It is clearly shown in Fig.
\ref{fig:channel_model} that the convergence happens whether the
AWGN or interference dominates. This is important for  the MLC
scheme because the interference to noise ratio varies for
different layers.

Although the optimality in Theorem \ref{thm:main} is established
for a large number of layers, next section will show how the power
of each layer can be allocated to minimize the number of required
layers.

\section{Practical design issues and numerical results}
\label{sec:design-issue}

MLC with multistage decoding can approach capacity if and only if
each lay uses a capacity achieving code whose rate is equal to the
layer's capacity \cite{wachsmannIT99}. The proposed MLC employs a
linear mapping, and we are free to allocate the power of each
layer. Consequently, the rate of each layer can be designed
flexibly. A good power allocation scheme is important in our case.
This section discusses these practical design issues and presents
numerical results. Two channels are used for simulation, a short
channel $h_1=\{1, 1\}$, and a randomly generated long channel
\[
h_2 = \{-0.432, -1.665, 0.125, 0.287, -1.146, 1.190, 1.189,
-0.037, 0.327, 0.174\}.
\]
\subsection{Achievable rate at each
layer}\label{sec:achievale-rate}

Since we are constrained in receiver structure,  we do not intend
to calculate the capacity of each layer. Instead, we will compute
its achievable rate under the given LMMSE equalizer. The
computation is based on the statistics of
 the bit probabilities computed by an actual receiver from a simulated channel output
sequence. No Gaussian assumption is made here. The achievable rate
of the layer $m$ sub-channel is given by
\begin{equation}\label{eqn:mth_layer_capacity}
    R_m = I\big(x_m(k); \widetilde{x}_m(k)\big) = 1 -
    \text{E}\big[-\log_2 P\big({x}_m(k) | \widetilde{x}_m(k)\big)\big]%H\big(x_m(k)\big) - H\big(x_m(k)|\widetilde{x}_m(k) \big)
\end{equation}
where   the {\it a posteriori} probability can be derived from
\eqref{eqn:likelihood} as
\begin{align}
P\big({x}_m(k) | \widetilde{x}_m(k)\big)  &=
\frac{P\big(\widetilde{x}_m(k) | {x}_m(k)\big)}
{P\big(\widetilde{x}_m(k) | +\!\sqrt{P_m}\big) +
P\big(\widetilde{x}_m(k) | -\!\sqrt{P_m}\big)},
\end{align}
and $\text{E}[-\log_2 P({x}_m(k) | \widetilde{x}_m(k))] $ is
evaluated using Monte Carlo simulation.  Thus, the overall
achievable rate of the multilevel scheme is $R_{MLC} =
\sum_{m=1}^MR_m$.

\subsection{Power allocation}\label{sec:power-allocation}

We consider three power allocation schemes as follows.

\begin{figure}[t]
 \centering
\includegraphics[width =5.5in]{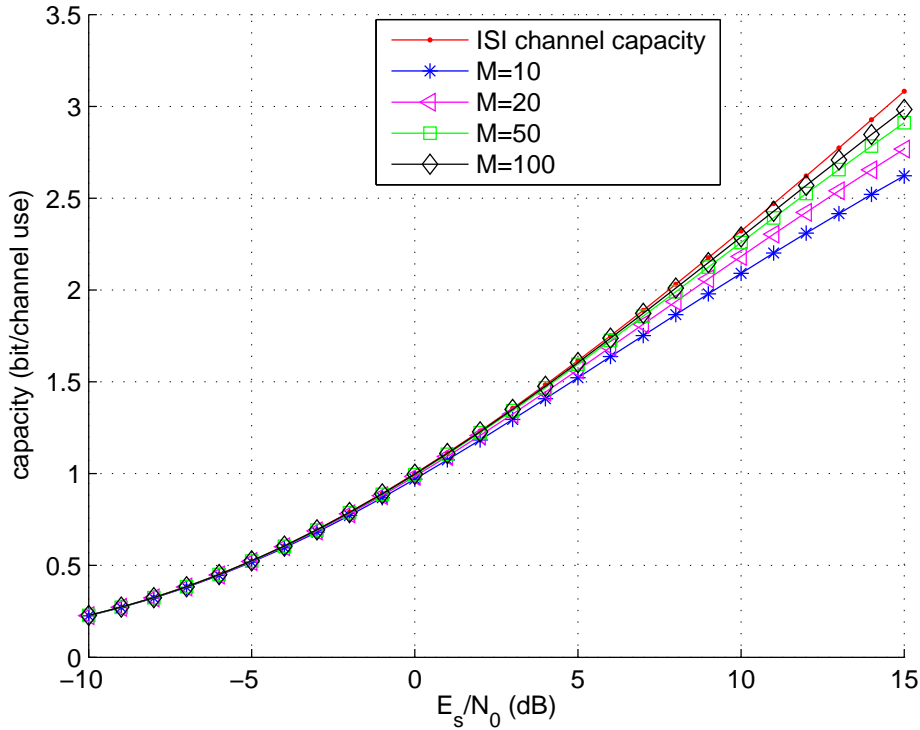}
      \caption{Achievable rate using equal power allocation for a short channel $h_1$}
      \label{fig:equal_pwr_alloc short}
\end{figure}

\begin{figure}[t]
 \centering
\includegraphics[width =5.5in]{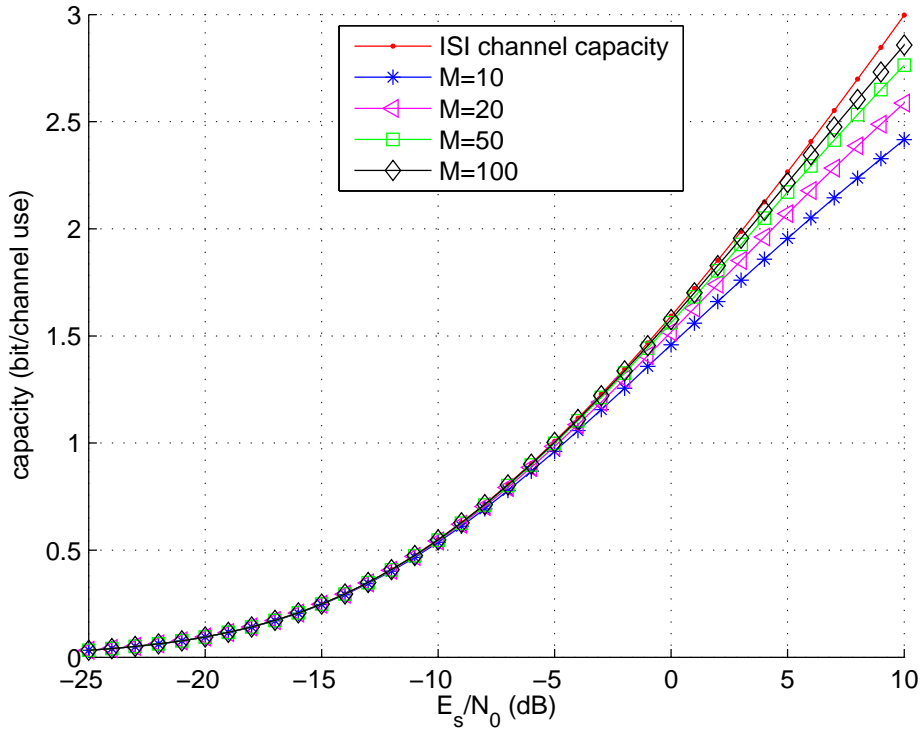}
      \caption{Achievable rate using equal power allocation for a long channel $h_2$}
      \label{fig:equal_pwr_alloc long}
\end{figure}

1) Equal power. In this scheme, each layer has the same power.
Fig. \ref{fig:equal_pwr_alloc short} and Fig.
\ref{fig:equal_pwr_alloc long} show the achievable rate of the
equal power MLC with $M = 10, 20, 50, 100$.

2) Equal distance. This power allocation scheme can produce a
constellation with equal distance. For example, to produce a
uniformly distributed $2^M$-ary ASK sequence, the power of each
layer should satisfy $P_j = 4P_{j+1}, \, j=1,\cdots,M-1$. This can
be extended to a QAM constellation if each dimension transmits an
independent ASK sequence.  Note, at moderate to high SNR, this
power allocation scheme is not capacity achieving since each layer
still operates at moderate SINR.

\begin{figure}[t]
\centering
\includegraphics[width =6.7in]{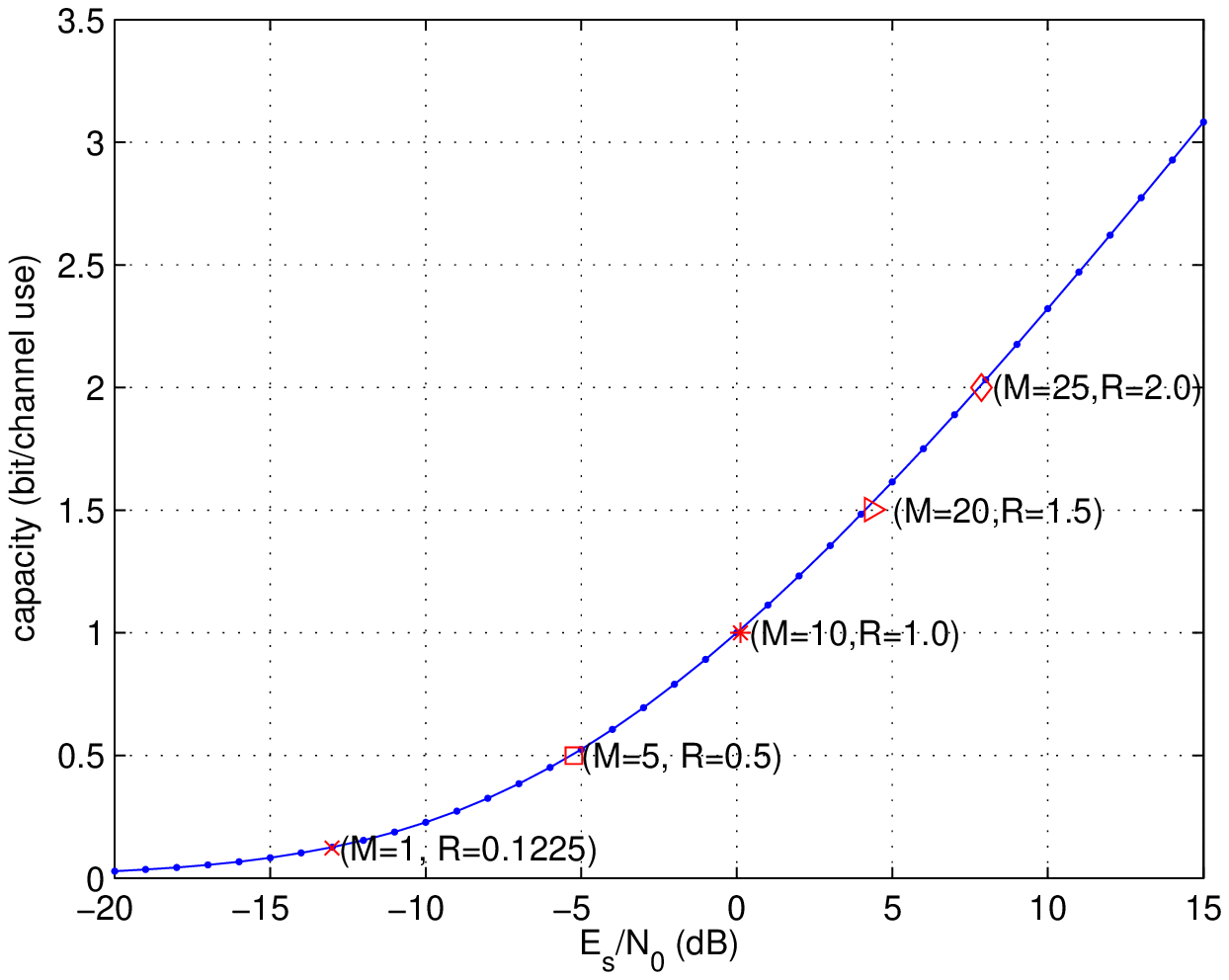}
      \caption{Achievable rate using equal rate power allocation for a short channel, where
$M$ is the number of layers and $R$ is the total achievable rate
of MLC. The solid curve is the ISI channel capacity with i.i.d.
Gaussian input.}
      \label{fig:equal_capacity short}
\end{figure}

\begin{figure}[t]
 \centering
\includegraphics[width =6.7in]{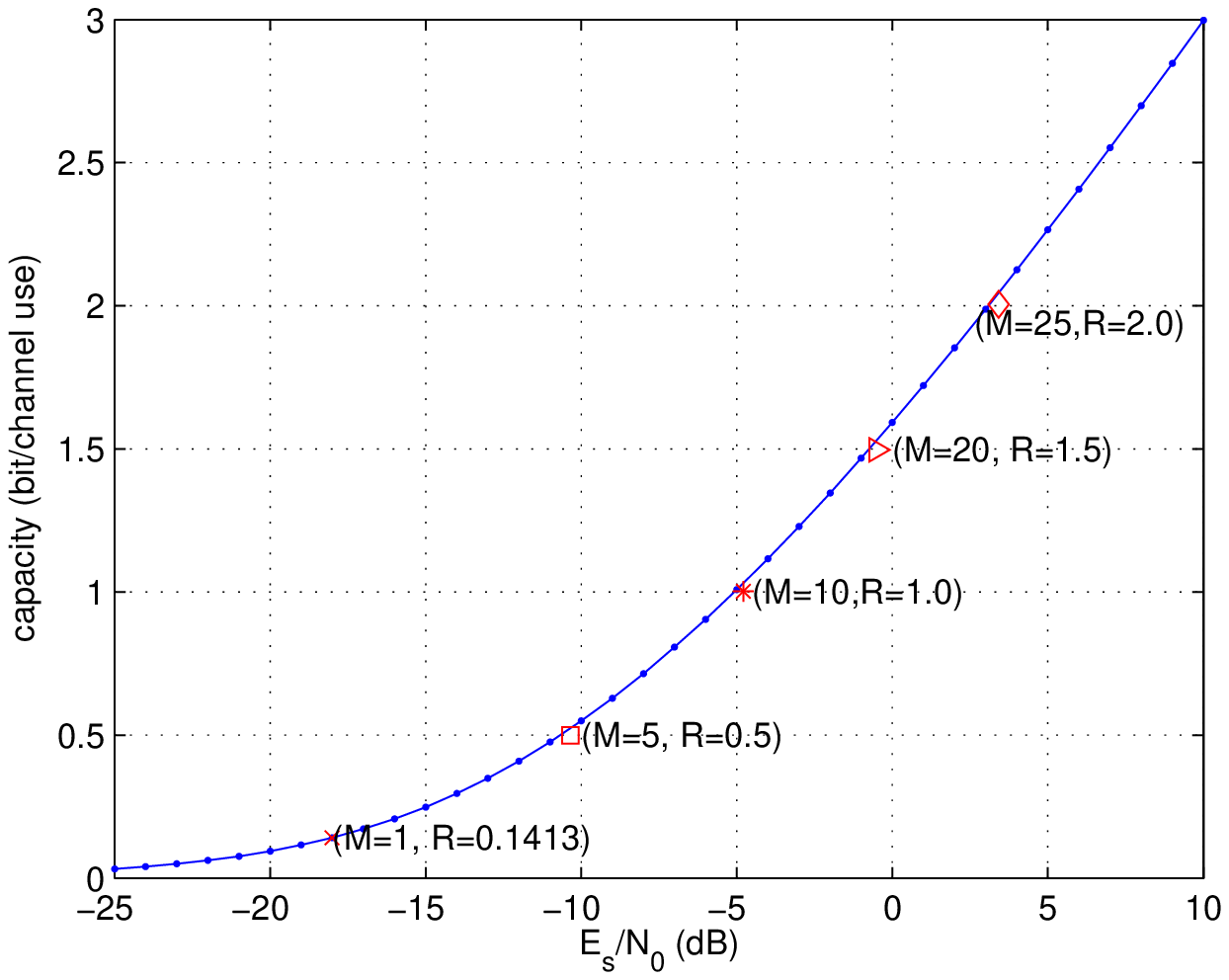}
      \caption{Achievable rate using equal rate power allocation for a long channel, where
$M$ is the number of layers and $R$ is the total achievable rate
of MLC. The solid curve is the ISI channel capacity with i.i.d.
Gaussian input.}
      \label{fig:equal_capacity long}
\end{figure}

3) Equal rate. This scheme allocates  the power in such a way that
the achievable rate at each layer is identical, i.e.,
$R_1=R_2=\cdots=R_M = R_{const}$, where $R_{const}$ is some fixed
rate. A simulation based numerical procedure is required to
determine the power of each layer, starting from layer $M$ to
layer $1$. Fig. \ref{fig:equal_capacity short} and Fig.
\ref{fig:equal_capacity long} compare the achievable rate of the
 equal rate MLC to the i.i.d.
Gaussian capacity for both a short and a long ISI channel.

In summary, both equal power and equal rate power allocation can
achieve channel capacity, while equal distance power allocation
can generate desired ASK or QAM constellations but does not
achieve capacity. Furthermore, equal rate power allocation
requires fewer layers than the equal power scheme to achieve the
same capacity.

\subsection{Code Design}\label{sec:coding}

A common tool for designing the component codes at each layer is
 equivalent channel model \cite{wachsmannIT99}, \cite{houIT03}. In
the proposed MLC with LMMSE filtering, the equivalent channel, as
seen in \eqref{eqn:eqv-chan}, is simply an AWGN channel with BPSK
input. Thus, many existing capacity achieving codes, such as
low-density parity-check (LDPC) codes \cite{gal-book}, and their
design technique, e.g., density evolution \cite{ rich01a}, can be
applied in a straightforward
 way.
The above code design process can be further simplified using
equal rate power allocation. In this case all equivalent channels
of all layers are equivalent to an AWGN channel with a given
capacity $R_{const}$. Therefore, we only need to design a single
code for this AWGN channel and apply the code and its
encoder/decoder pair to all layers. Note, a random interleaver
will be used for each layer to avoid decision feedback error
propagation in this case.

The second task  is to assign a code rate for each layer. When a
capacity achieving code with infinite block length is available,
we can use  capacity rule \cite{wachsmannIT99} so that the code
rate at layer $j$ is equal to the achievable rate $R_j$. When code
is not capacity achieving,  the code rate must be smaller than the
achievable rate. Furthermore, since the MLC scheme relies on
perfect decoding of all layers, the code rate must be designed so
that all layers will have a low BER at a given overall SNR
simultaneously. Other techniques such as random coding bound
\cite{wachsmannIT99} can also be applied to design the rate.

\section{Extension and conclusion}

This paper proposes an MLC and LMMSE equalization scheme and shows
that it can achieve channel capacity for ISI channels with i.i.d
Gaussian inputs. A further extension to achieve the ultimate
``water-filling'' capacity is also straightforward by
incorporating a spectral shaping filter at the transmitter and a
corresponding LMMSE equalizer at the receiver. The scheme is
computationally efficient and is especially attractive for systems
with severe ISI.

The principle  of MLC with linear mapping and multistage decoding
can be applied to correlated or block fading channels with unknown
channel states. The fundamental idea is using the decoded layers
as training symbols. The receiver then estimates the channel based
on the training symbols.  As long as the power of each layer is
very small, estimation and decoding can be decoupled without loss.
This is under current investigation.

\bibliographystyle{IEEEbib}
\bibliography{MLC}

\end{document}